# Coherent spin dynamics in gadolinium-doped CaWO$_4$ crystal

E. I. Baibekov[1,*], M. R. Gafurov[1], D. G. Zverev[1], I. N. Kurkin[1], A. A. Rodionov[1],

B.Z. Malkin[1], B. Barbara[2]

[1] *Kazan Fedral University, 18 Kremlyovskaya Street, Kazan 420008, Russia*

[2] *Institut Néel, CNRS, BP166, 38042 Grenoble Cedex 9*

*and Université Joseph Fourier, France*

*Corresponding author. E-mail address: edbaibek@gmail.com

**Abstract**

We report the first observation of Rabi oscillations in the spin-7/2 ensemble of trivalent gadolinium ions hosted in CaWO$_4$ single crystal. A number of transitions within the lowest electronic multiplet $^8S_{7/2}$ of Gd$^{3+}$ ion are studied using a combination of continuous-wave and pulsed electron paramagnetic resonance spectroscopy. The corresponding Rabi damping curves and the spin coherence times are detected at varying strengths of the microwave field. These data are well reproduced by a theoretical model which accounts for the intrinsic inhomogeneity of the microwave field within the microwave resonator and the magnetic dipole interactions in the diluted spin ensemble. The results indicate that the studied 8-level ground manifold of Gd$^{3+}$ ion can represent an effective three-qubit quantum system.

## I. INTRODUCTION

The concept of bulk quantum computing (QC) implies that the same computation algorithm is executed simultaneously within a large number of identical noninteracting spin systems [1]. The spin states are addressed using conventional methods of pulsed magnetic resonance spectroscopy. A simple transient pulse with frequency matching that of a certain quantum transition can be viewed as one-qubit gate [2]. For example, a $\pi$ pulse flipping the spin-1/2 states represents quantum NOT gate. One clear advantage of this scheme is that it provides the grounds for testing the basic concepts of QC with no specific instrumentation apart from commercial nuclear magnetic resonance (NMR) or electron paramagnetic resonance (EPR) spectrometer. In particular, implementation of 7-qubit Shor's factoring algorithm using the nuclear spin states of perfluorobutadienyl iron complex culminated the first rapid progress of NMR quantum computation in the early 2000s [3]. A thorough theoretical analysis, however, showed that the sole use of nuclear spin states with achievable population difference $< 10^{-3}$ actually provided no quantum entanglement and restrained the scalability of the NMR quantum computer [4,5]. On the contrary, typical X-band EPR provides sufficient pure component of an electron spin density matrix at temperatures ~ 1 K to overcome aforementioned issues. In the case of EPR QC, the following problems arise:

(i) The electron spin coherence times are usually much shorter than the nuclear ones. The number of coherent QC operations is much smaller, even considering the difference in the excitation frequency.

(ii) The possibility to excite different transitions during the NMR pulse sequence was a key capability that induced the first success of multiqubit NMR QC. Up to the moment, no commercialized EPR instrumentation allows complete control over the pulse shape and frequency, with a possible exception of Bruker SpinJet-AWG system [6]. Two-qubit electronic or hybrid electron-nuclear systems can be operated by pulsed electron-electron (ELDOR) or electron-nuclear (ENDOR) double resonance techniques, respectively [7].

The best-known implementations of solid-state EPR QC include: NV-centers in diamond [8], single electron spin in a quantum dot [9], single-molecule magnets [10-12], organic radicals [7], rare earth and transition metal ions in crystals [13,14]. Most of these quantum systems have rich energy level structure, although large difference in the excitation frequencies complicates the selective excitation of various quantum transitions during the pulse sequence. It is well-known that the lowest states of ions with half-filled valence shell, so-called S-state ions ($Fe^{3+}$, $Mn^{2+}$, $Gd^{3+}$, $Eu^{2+}$, ...) are characterized by certain total spin $S$ and zero total orbital moment, so that spin-lattice and crystal-field interactions are suppressed to a certain extent [15]. These ions can be actually treated as particles bearing large spin $S$. Zeeman interaction creates nearly equidistant lower energy level scheme. Rabi oscillations (ROs) of $Mn^{2+}$ ions in MgO crystal and of $Fe^{3+}$ in ZnO (both with $S = 5/2$) have been studied recently [16-18]. S-state lanthanide ions ($Eu^{2+}$, $Gd^{3+}$, $Tb^{4+}$) have more complex energy level scheme and larger number of observable transitions due to higher spin $S = 7/2$. Until recently, it seems that no demonstration of coherent spin dynamics of the S-state lanthanide ions has been published [19]. In the present article, we provide the experimental evidence of collective spin nutations in the ensemble of trivalent gadolinium ions incorporated in $CaWO_4$ single crystal.

## II. EXPERIMENTAL PROCEDURE

The sample of $CaWO_4$:$Gd^{3+}$ single crystal was grown by Czochralski method in Magnetic Resonance Laboratory of Kazan Federal University. Nominal concentration of gadolinium ions equaled 0.01 at. % ($C = 1.28 \cdot 10^{18}$ ions per cc.), which was verified within relative error of 20% by comparative measurements of continuous-wave (cw) EPR spectral intensities with respect to the reference sample of $LiYF_4$:$Gd^{3+}$ with known concentration of 0.5 at. %. In order to reduce the effect of microwave (mw) field inhomogeneity on the Rabi decay times, a small plate with approximate dimensions $3 \times 2 \times 0.5$ mm was cut from the original sample. A quartz tube containing this smaller sample was placed close to the center of standard cylindrical dielectric resonator Bruker EN 4118X-MD4 of X-band Elexsys 580/680 EPR spectrometer. Parallel orientation of the crystal sample axis $c$ with respect to the static magnetic field $\boldsymbol{B}_0$ (see Fig. 1) was checked by comparison of the calculated and measured cw EPR spectra.

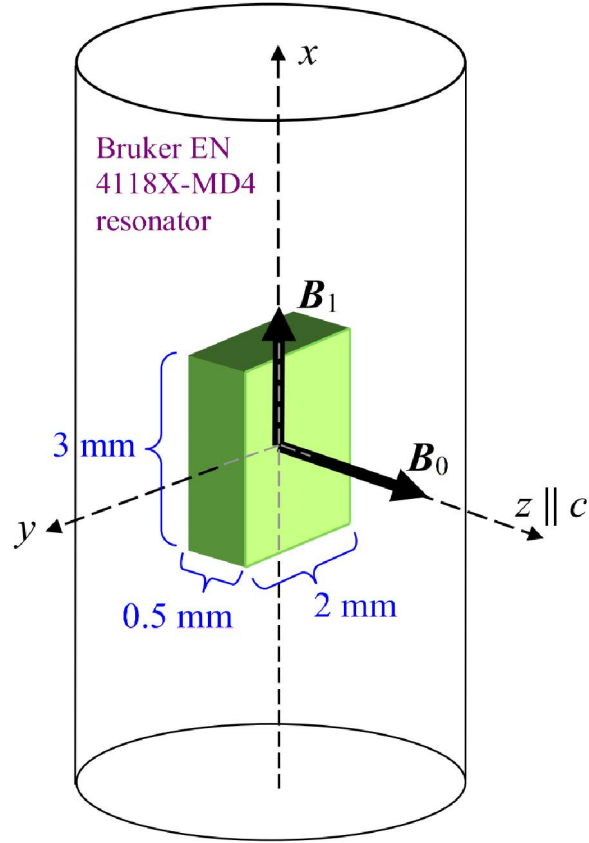

FIG. 1. A scheme illustrating the orientation of the crystal sample of $CaWO_4$:$Gd^{3+}$ with respect to the resonator axis, static and mw field vectors $\boldsymbol{B}_0$ and $\boldsymbol{B}_1$.

Room-temperature cw EPR spectra were acquired at $\omega_0/2\pi = 9.66$ GHz and in the field range of $\boldsymbol{B}_0 = 0\text{-}10^4$ G. Pulsed measurements were accomplished in the temperature range $T = 6\text{-}150$ K. Spin-lattice relaxation times $T_1$ and phase memory times $T_2$ were obtained by means of inversion-recovery ($\pi - \tau - \pi/2 - \tau' - \pi$) and Hahn spin echo ($\pi/2 - \tau - \pi$) pulse sequences, respectively, where $\tau$ was incremented and $\tau'$ kept constant. The durations of $\pi/2$ and $\pi$ pulses equaled 8 and 16 ns, respectively. Each data point of ROs was obtained as a result of the pulse sequence shown in Fig. 2, where the transient pulse of length $t$ was followed by the spin-echo detection sequence. After that, the sequence was repeated with time $t$ incremented by 4 ns. As a result, we obtained time dependence of the longitudinal magnetic moment $M_Z(t)$ of the Gd$^{3+}$ ($S = 7/2$) spin ensemble in the form of the decaying oscillations. Various Rabi frequencies $\Omega_R$ were obtained by changing the mw field attenuation.

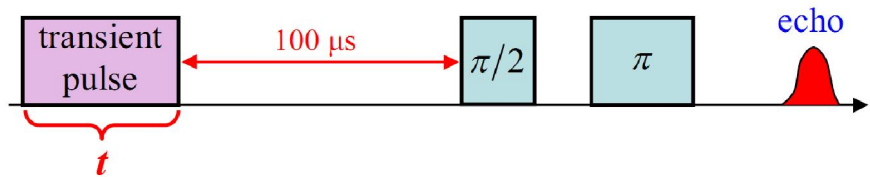

FIG. 2. Transient pulse sequence that was applied in order to record ROs of Gd$^{3+}$ ions in CaWO$_4$ crystal. The duration of $\pi/2$ and $\pi$ pulses equaled 8 and 16 ns, respectively. The time $t$ was varied from 0 to 2 μs with a step of 4 ns.

A tiny coal sample was used to calibrate the distribution of the mw field $\boldsymbol{B}_1(x)$ along the resonator axis. The relative amplitudes $B_1(x)/B_1^{(\max)}$ were determined by comparing the cw EPR signals at different positions of a sample tube with respect to the field antinode of the resonator (Fig. 3).

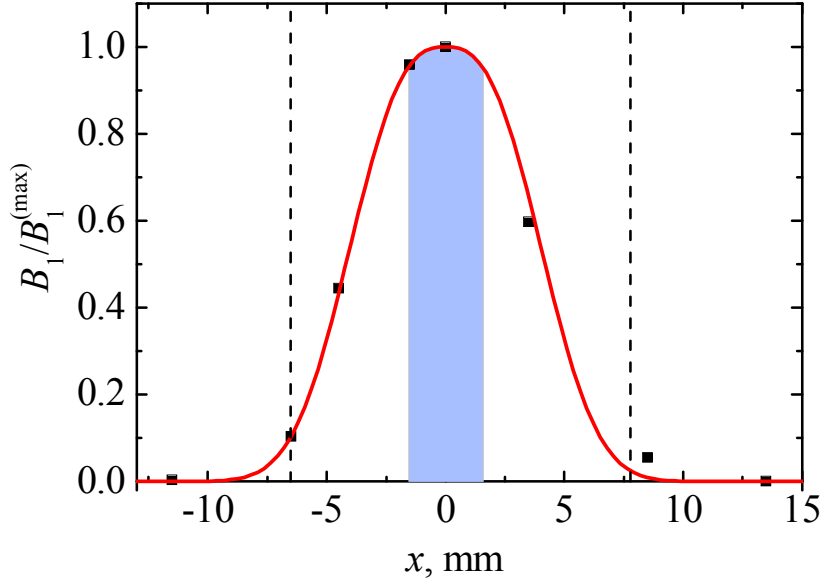

FIG. 3. The distribution of the mw field amplitude along the resonator axis. Experimental data and their fit by the generalized Gaussian distribution function $\exp\left[-(x/a)^b\right]$ ($a = 4.8$ mm, $b = 2.7$) are represented by the black squares and the red line, respectively. A blue fill area corresponds to the approximate position occupied by the crystal sample of $CaWO_4:Gd^{3+}$. Vertical dashed lines show the borders of the 13 mm-long resonator.

## III. RESULTS AND DISCUSSION

### A. EPR spectra and transitions

The cw EPR spectra of $Gd^{3+}$ ions in various host crystals [20,21], particularly in $CaWO_4$ single crystal [22-26], had been extensively studied. The present-work cw EPR spectra recorded at $\boldsymbol{B}_0 \parallel c$ are shown in the top part of Fig. 4. There, ca. 17 electron spin transitions $M_1 \leftrightarrow M_2$ ($M_i = -7/2...7/2$) were identified by means of the numerical diagonalization of the spin Hamiltonian. Due to weak mixing of the ground $^8S_{7/2}$ multiplet with excited states, the Hamiltonian contained crystal-field terms, in addition to Zeeman interaction:

$$H = g\mu_B S_Z B_0 + B_2^0 O_2^0 + B_4^0 O_4^0 + B_4^4 O_4^4 + B_6^0 O_6^0 + B_6^4 O_6^4, \quad (1)$$

where $O_p^k$ are Stevens operators constructed on the basis of 8 spin-7/2 states, $\mu_B$ is Bohr magneton. The g-factors $g_\parallel = 1.991$, $g_\perp = 1.992$ and the crystal field parameters $B_2^0 = -0.0298$, $B_4^0 = -3.80 \cdot 10^{-5}$, $B_4^4 = -2.34 \cdot 10^{-4}$, $B_6^0 = 1.98 \cdot 10^{-8}$, $B_6^4 = 1.59 \cdot 10^{-8}$ (in units of cm$^{-1}$, at 293 K), as well as their temperature variations, had been determined previously [23]. In our calculations, we used an isotropic g-factor $g = 1.99$ for simplicity. The observed magnitudes of $B_0$ corresponding to the studied EPR transitions varied with temperature due to temperature dependence of the crystal-field parameters (especially, $B_2^0$). For the high-field $7/2 \leftrightarrow 5/2$ transition, $B_0$ increased from ~ 9700 G to 9900 G upon decreasing the temperature from 300 K to 6 K. The calculated energies of the $^8S_{7/2}$ electronic levels versus magnetic field are presented on the bottom part of Fig. 4. Numerically calculated EPR spectra (not shown) were in good agreement with the experimental data. The observed transitions were classified as follows:

(i) The most intense were $M \leftrightarrow M \pm 1$ transitions excited by the interaction with the mw field $H_{MW} = 2gS_X B_1 \cos \omega_0 t$ (shown in Fig. 4 by green arrows).

(ii) The tetragonal-symmetry crystal field interaction mixed $M$ and $M \pm 4$ states, which resulted in lower-intensity $M \leftrightarrow M \pm 3$ and $M \leftrightarrow M \pm 5$ transitions (blue arrows).

(iii) The smallest signals came from the forbidden transitions $M \leftrightarrow M \pm 2$ and $M \leftrightarrow M \pm 4$ (red arrows). The necessary mixing of the states $M \leftrightarrow M \pm 1$, $M \leftrightarrow M \pm 3$ presumably originated from the presence of nearby crystal lattice defects lowering the tetragonal symmetry, and/or hyperfine interactions with $^{155}$Gd and $^{157}$Gd nuclei.

One can identify at least four field regions (near 660, 1780, 2670 and 5000 G, see brown rectangles on the lower part of Fig. 4), where two and more transitions connecting at least three energy levels fall into the frequency band of $9.6 \pm 0.5$ GHz accessible by the pulsed ELDOR techniques. For instance, at $B_0 = 2670$ G, a subgroup of four electronic states can be manipulated simultaneously through the transitions $7/2 \leftrightarrow -3/2$, $7/2 \leftrightarrow 1/2$ and $7/2 \leftrightarrow 5/2$ within the frequency bandwidth of ~ 1 GHz.

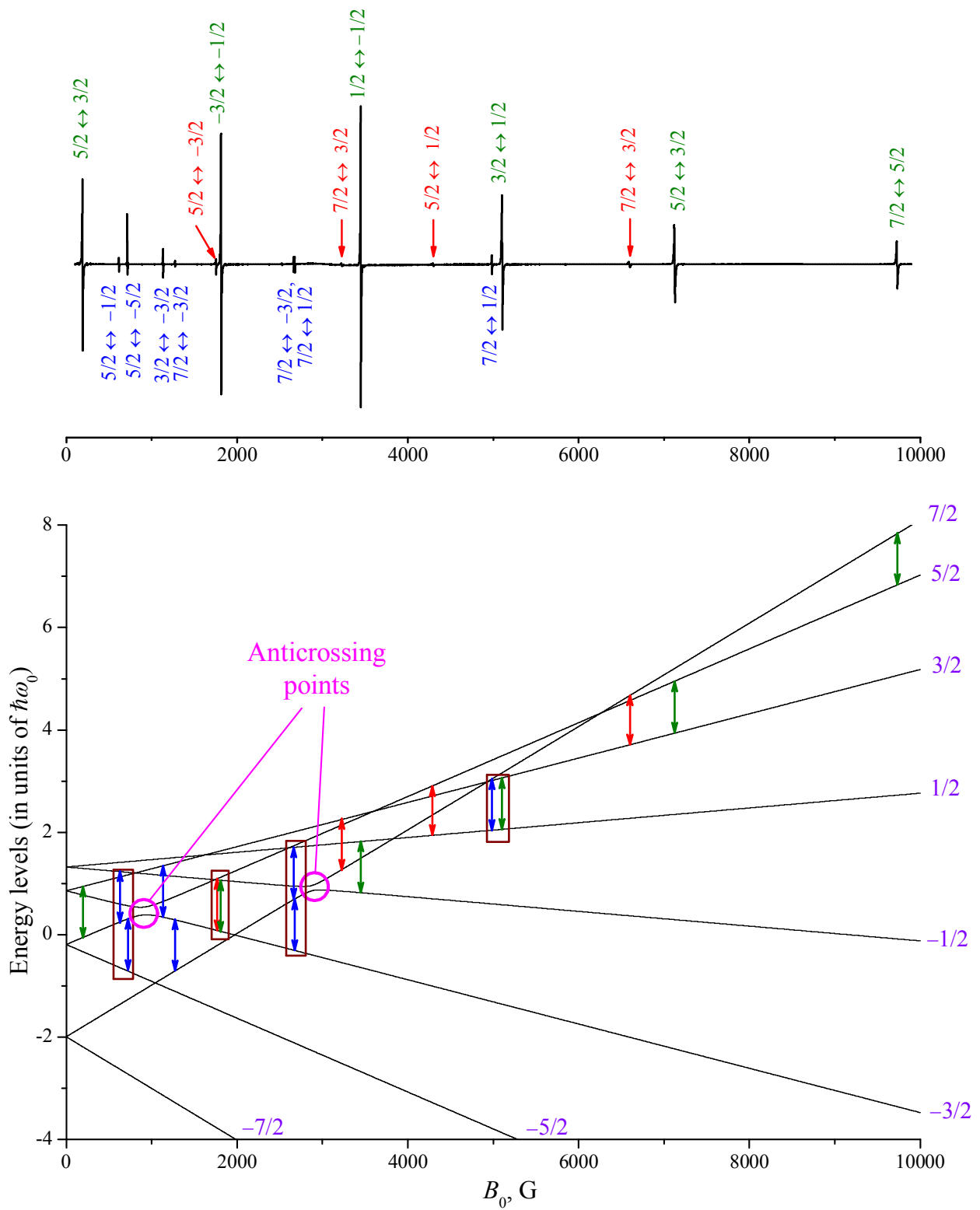

FIG. 4. X-band EPR spectra of $Gd^{3+}$ ion in $CaWO_4$ crystal (top), and the calculated energy levels of the lowest $^8S_{7/2}$ multiplet versus magnetic field (bottom). $T = 300$ K, $\omega_0/2\pi = 9.66$ GHz. Different transitions are marked with green, blue and red arrows, see text. Two anticrossing points that arise from mixing of $5/2 \leftrightarrow -3/2$ and $7/2 \leftrightarrow -1/2$ spin levels by the $B_4^4 O_4^4$, $B_6^4 O_6^4$ terms of the crystal field interaction are marked with pink circles. Brown rectangles show regions containing multiple transitions accessible by the pulsed ELDOR.

Due to narrower EPR lines at low temperatures, an observation of the hyperfine structure was sometimes possible. Field-swept echo-detected spectrum near the $1/2 \leftrightarrow -1/2$ electronic transition with full width at half-maximum (FWHM) ~ 2.5 G at 15 K revealed the presence of $^{155}$Gd and $^{157}$Gd isotopes with the nuclear spin $I = 3/2$ (natural abundance 14.7 % and 15.7 %, respectively), see Fig. 5. The positions of the hyperfine components agree well with the previously reported hyperfine constants $A_{155} = 12.40$ and $A_{157} = 16.28$ MHz [22].

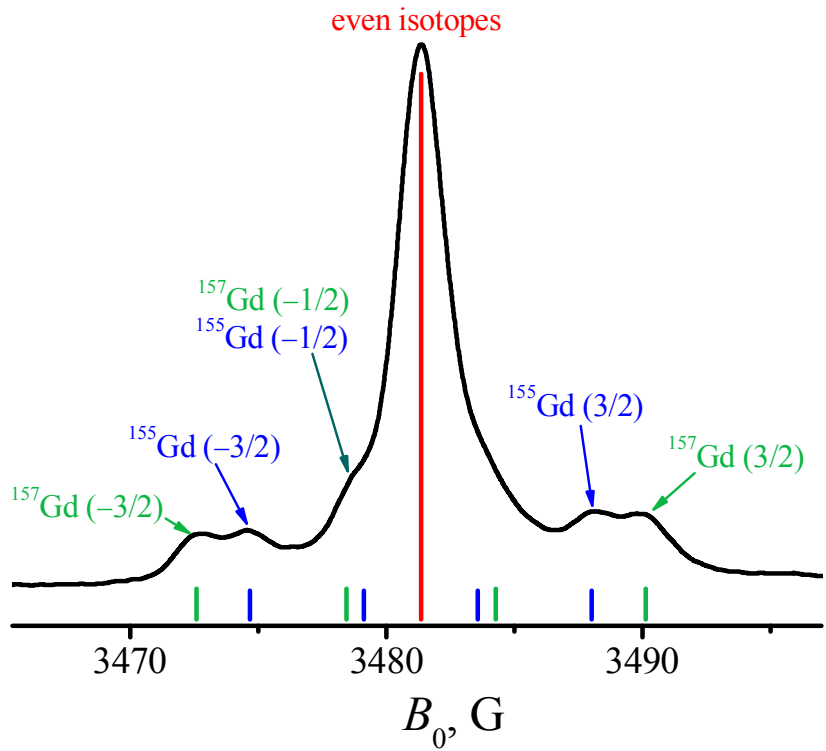

FIG. 5. Field-swept echo-detected EPR spectrum near the $1/2 \leftrightarrow -1/2$ electronic transition revealing the hyperfine structure due to the presence of $^{155}$Gd and $^{157}$Gd isotopes (nuclear spin 3/2). The bars show the calculated positions of the hyperfine lines with respect to the central line and their relative intensities. $T = 15$ K, $\omega_0/2\pi = 9.75$ GHz.

### B. Phase memory and spin-lattice relaxation times

The $T_2$ and $T_1$ times corresponding to the intense $M \leftrightarrow M \pm 1$ transitions were first measured at 6 and 15 K. The results are presented in the last two columns of Table I. Our experimental $T_1$ values ~ 8 ms at 6 K and ~ 1 ms at 15 K are comparable to the previously reported data acquired by the pulse saturation technique [27]. The corresponding $T_2$ times of ~ 5÷10 μs depended on temperature, which indicated the possible impact of the spectral diffusion process associated with the spin-lattice relaxation of the surrounding spins [28]. Nearly two times longer $T_2$ values in the case when the magnetic field $B_0 = 3473$ G was tuned to the $m_I = -3/2$ hyperfine component of the $^{157}$Gd $1/2 \leftrightarrow -1/2$ transition suggested that the dominant contribution came from the instantaneous diffusion process [29]. This suggestion was also corroborated by the observation of sufficiently longer $T_2$ times upon 10-time increase of the $\pi/2$ and $\pi$ pulse duration. Due to negligible amount of the magnetic $^{43}$Ca and $^{183}$W isotopes, and the gadolinium dilution sufficient enough to rule out the exchange interactions, the experimental $T_2$ values were simulated taking into account the magnetic dipole interactions between the $Gd^{3+}$ ions:

$$1/T_2^{(Th)} = \Gamma_{ID} + \Gamma_{SD}. \quad (2)$$

Above, $\Gamma_{ID}$ is the rate of the instantaneous diffusion process associated with the change of the local magnetic field induced by the flipping nearby gadolinium spins at the application of the second ($\pi$) pulse of the spin-echo sequence [29]. This rate was estimated as

$$\Gamma_{ID} = \gamma_T \Delta \omega_d^{(1/2)} \left\langle \sin^2 \frac{\theta_\pi}{2} \right\rangle, \quad (3)$$

where $\Delta \omega_d^{(1/2)} = 4\pi^2 g^2 \mu_B^2 C / 3\sqrt{3}\hbar = 1.05 \cdot 10^6$ s$^{-1}$ is the dipolar half-width of the EPR line calculated for the diluted ensemble of the spin-1/2 particles [28,30] (for the $M \leftrightarrow M+1$ transition, $\pi$ pulse changed the spin projection by unity, thus the change of the local field was equivalent to the one in the spin-1/2 ensemble); $\theta_\pi$ is the angle by which the $\pi$ pulse rotates a given spin packet, $\langle ... \rangle$ denotes averaging over all spin packets within the EPR line of spectral density $g(\omega)$; $\langle \sin^2 \theta_\pi/2 \rangle$ was calculated according to the Ref. [29] as

$$\left\langle \sin^2 \frac{\theta_\pi}{2} \right\rangle = \int d\omega g(\omega) \frac{\Omega_R^2}{(\omega - \omega_0)^2 + \Omega_R^2} \sin^2 \left\{ \frac{\pi}{2} \left[ \frac{(\omega - \omega_0)^2}{\Omega_R^2} + 1 \right] \right\}. \quad (4)$$

The Rabi frequency $\Omega_R/2\pi = 1/(2t_\pi)$ equaled 31.2 MHz in the case of the $t_\pi = 16$ ns $\pi$-pulse duration. The above integral was calculated numerically for all studied transitions, where $g(\omega)$ were acquired directly from the field-swept echo-detected EPR spectra, which showed combinations of nearly Lorentzian lines $\sim \left( (\omega - \omega_i)^2 + \sigma^2 \right)^{-1}$ with FWHM ($2\sigma$) varying in the range of $\sim 2.5 \div 15$ G (see Fig. 5 and the 3$^{rd}$ column of the Table I). At $t_\pi = 16$ ns, we obtained $\langle \sin^2 \theta_\pi/2 \rangle$ $\sim 0.7$-$0.8$, meaning that the most part of the spin packets were excited by the spin-echo sequence. At $t_\pi = 160$ ns and for the hyperfine $^{157}$Gd transition, $\langle \sin^2 \theta_\pi/2 \rangle \sim 0.04$, which explained longer $T_2 \sim$ 25 µs.

The population factor $\gamma_T$ was added into Eq. (3) to account for only those spins that populate the $M$ and $M+1$ levels involved in the transition with the corresponding energies $E_M$ and $E_{M+1}$. Assuming that the spin ensemble was initially in the state of temperature equilibrium,

$$\gamma_T = \frac{\exp(-E_M/k_B T) + \exp(-E_{M+1}/k_B T)}{\sum_{M'=-S}^{S} \exp(-E_{M'}/k_B T)}. \quad (5)$$

At infinite temperature, $\gamma_\infty = 0.25$. It can differ considerably from 0.25 at low temperature, giving longer $T_2$ times $\sim 10$ μs for the high-field $5/2 \leftrightarrow 3/2$ and $7/2 \leftrightarrow 5/2$ transitions; in particular, $\gamma_T = 0.124$ for the $7/2 \leftrightarrow 5/2$ transition at 6 K.

The second studied mechanism was the spectral diffusion [28,29], in which the local fields were fluctuated by spontaneous flips of the nearby spins induced by the spin-lattice and by the magnetic dipole interactions. The latter, however, were negligible since the dipolar-induced flip-flop processes are unfavorable when $\Delta\omega_d^{(1/2)} \ll$ FWHM, i.e. when the lines are inhomogeneously broadened. The rate of the spectral diffusion was calculated as follows

$$\Gamma_{SD} = \sqrt{f_S \Delta\omega_d^{(1/2)} \tilde{T}_1^{-1}/2}, \quad (6)$$

where we added the factor $f_S \sim 3$ to $\Delta\omega_d^{(1/2)}$ in order to account for the random $M' \rightarrow M''$ flips with $|M'-M''| > 1$ ($f_S = 1$ for $S = 1/2$), and $\tilde{T}_1^{-1}$ was the flip rate averaged over all transitions at a given temperature (on the basis of our experimental $T_1$ values, we chose $\tilde{T}_1 = 8$ ms at 6 K and $\tilde{T}_1 = 1$ ms at 15 K). According to our calculations, instantaneous diffusion prevailed over the spectral diffusion. Only in the case of longer pulses and/or hyperfine transitions the contribution of the latter was significant. One can see in Table I that the calculated $T_2^{(Th)}$ are in reasonable agreement with the experimental $T_2$ values.

TABLE I. The phase memory ($T_2$) and the spin-lattice relaxation ($T_1$) times measured at 6 and 15 K for various quantum transitions. $\omega_0/2\pi = 9.75$ GHz. The calculated phase memory time $T_2^{(\text{Th})}$, see Eq. (2), contained the contributions from the instantaneous ($\Gamma_{\text{ID}}$) and spectral ($\Gamma_{\text{SD}}$) diffusion mechanisms.

| $B_0$, G * (transition) | FWHM ($2\sigma$), G | $T$, K | $\Gamma_{\text{ID}}$, $10^5$ s$^{-1}$ | $\Gamma_{\text{SD}}$, $10^5$ s$^{-1}$ | $T_2^{(\text{Th})}$, μs | $T_2$, μs | $T_1$, ms |
|---|---|---|---|---|---|---|---|
| 1806 ($-3/2 \leftrightarrow -1/2$) | 2.7 | 6 | 1.8 | 0.14 | 5.1 | 5.4 | 8 |
| | | 15 | 1.9 | 0.40 | 4.3 | 5.1 | 0.97 |
| 3483 ($1/2 \leftrightarrow -1/2$) | 2.5 | 6 | 1.9<br>1.1†<br>0.69§<br>0.09†§ | 0.14<br>0.14†<br>0.14§<br>0.14†§ | 4.8<br>7.8<br>12<br>44 | 6.5<br>10.2†<br>12§<br>25†§ | 8<br>6†<br>2.6§<br>3.5†§ |
| | | 15 | 2.1<br>1.2†<br>0.09†§ | 0.40<br>0.40†<br>0.40†§ | 4.0<br>6.1<br>20 | 5.4<br>8.9†<br>13.1†§ | 1.2<br>1.15† |
| 5178 ($3/2 \leftrightarrow 1/2$) | 10 | 6 | 1.4 | 0.14 | 6.4 | 6.4 | 7 |
| | | 15 | 1.6 | 0.40 | 4.8 | 5.3 | 0.9 |
| 7240 ($5/2 \leftrightarrow 3/2$) | 13 | 6 | 1.2 | 0.14 | 7.4 | 8.6 | 7 |
| | | 15 | 1.6 | 0.40 | 5.0 | 5.4 | 0.8 |
| 9892 ($7/2 \leftrightarrow 5/2$) | 15 | 6 | 0.94 | 0.14 | 9.2 | 13.1 | 10 |
| | | 15 | 1.5 | 0.40 | 5.2 | 10.6 | 0.85 |

\* at 6 K
† the data are presented for $m_I = -3/2$ transition of the $^{157}$Gd isotope ($B_0 = 3473$ G)
§ the durations of $\pi/2$ and $\pi$ pulses were 10 times longer (80 and 160 ns, respectively)

In order to trace the temperature variation of the relaxation times, the dependences $T_1(T)$ and $T_2(T)$ corresponding to the high-field transition ($7/2 \leftrightarrow 5/2$) were acquired in the range of 6-150 K, see Fig. 6. The temperature variation of the spin-lattice relaxation rate was simulated by the following expression

$$T_1^{-1}(T) = AT + BT^n, \quad (7)$$

where the first term corresponded to the direct process, and the second term – to Raman-type relaxation process [31]. Our data were fitted by the Eq. (7) with $A = 8$ K$^{-1}$s$^{-1}$, $B = 0.02$ K$^{-4}$s$^{-1}$, and $n = 4$. Above 10 K, the dominant contribution $\sim T^4$ came from the Raman-type process. The index $n = 4$ differed slightly from $n = 3$ and 3.6 that had been obtained previously [25]. According to the existing models of the spin-lattice relaxation of an S-state ion [31] and of the paramagnetic center situated at the defect site [32], $n = 3 \div 5$.

The phase relaxation rate $T_2^{-1}(T)$ was calculated using the Eq. (2), where the term $1/T_1$ associated with the spin-lattice interaction was added at higher temperatures. Our calculations showed three temperature intervals: (a) below 30 K, (b) 30-80 K, and (c) above 100 K, where different processes played a key role in the phase relaxation. The instantaneous diffusion, which was temperature-independent, dominated the relaxation in the first interval, while the spectral diffusion and the $1/T_1$ term dominated in the second and third intervals, respectively.

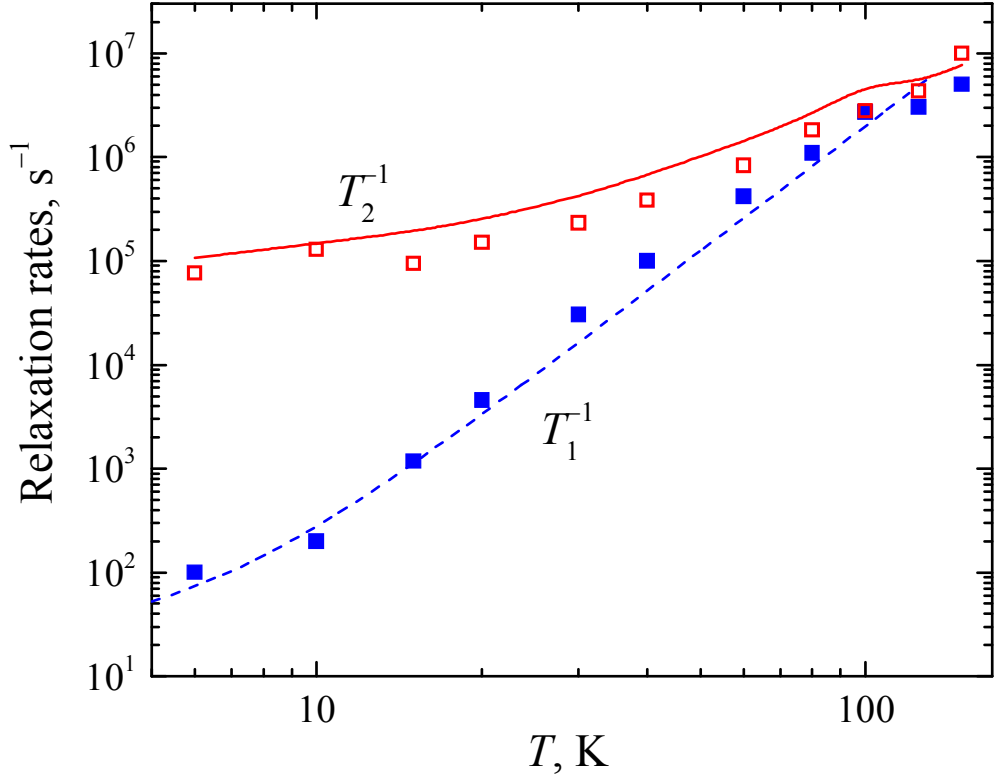

FIG. 6. Phase ($T_2^{-1}$) and spin-lattice ($T_1^{-1}$) relaxation rates in the temperature range 6-150 K. Experimental data are represented by symbols. Solid and dashed lines correspond to the calculations by Eq. (2) with the term $T_1^{-1}$ added, and the fit to Eq. (7), respectively.

### C. Rabi oscillations

ROs corresponding to all studied $M \leftrightarrow M \pm 1$ transitions were acquired in the presence of resonant mw field. Rabi frequencies $\Omega_R/2\pi \sim$ 0.97÷33 MHz were determined from the observed spin nutation signal. Up to 100 Rabi half-periods were observed, each corresponding to one-qubit NOT gate. The Rabi time $\tau_R$ (damping time of ROs) was roughly proportional to $\Omega_R^{-1}$. Firstly, ROs of various transitions were measured at constant mw field attenuation, some of them are presented in Figs. 7 and 8. The resulting decay profiles were found almost independent of the transition field $B_0$, gadolinium isotope and temperature. The damping was essentially slower than exponential. These facts ruled out the dipolar interactions as a major source of the Rabi damping, and suggested that the greatest contribution came from the distribution of the mw field $B_1(r)$ within the sample volume.

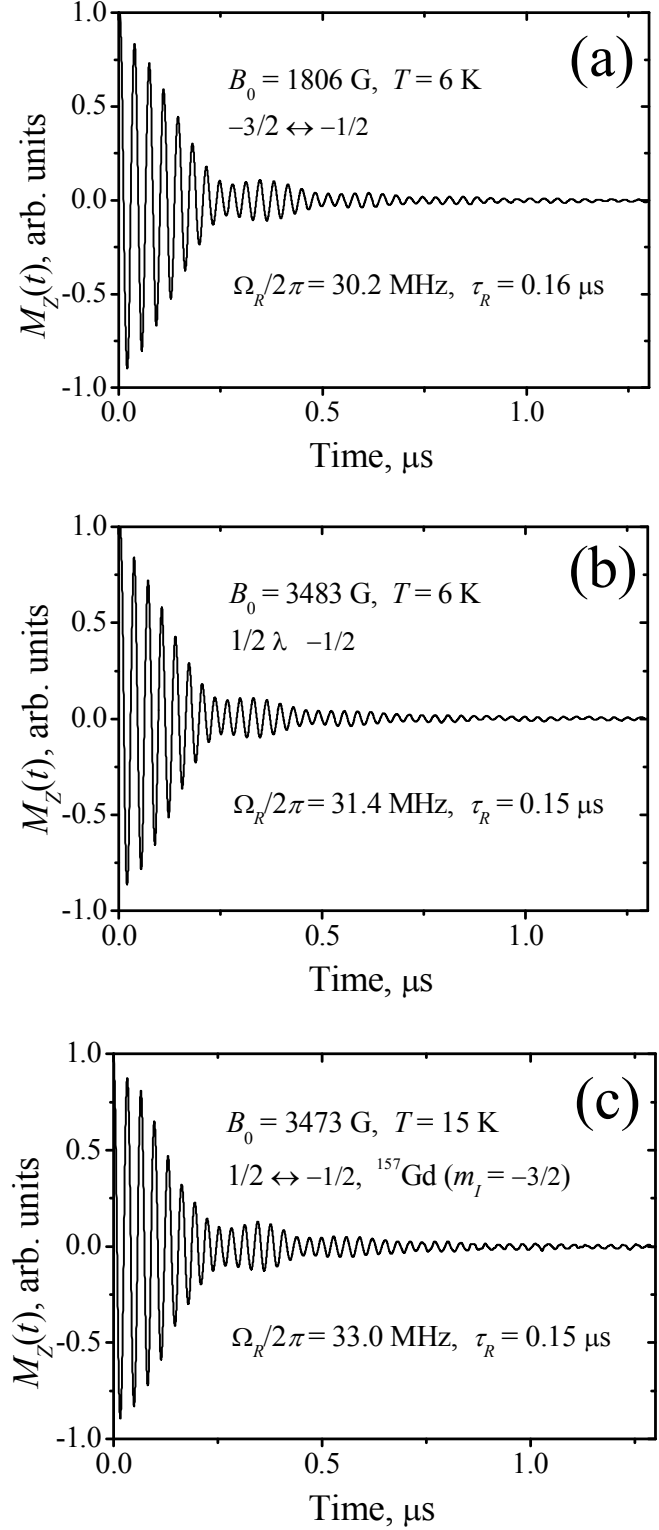

FIG. 7. Rabi oscillations between the quantum states of the spin-7/2 trivalent gadolinium ion in CaWO4 crystal. The level notations are the same as in Figs. 4 and 5. Rabi frequencies $\Omega_R/2\pi$ and Rabi times $\tau_R$ were determined from these experimental data as the frequencies of the damped oscillations and the characteristic damping times, respectively. $\omega_0/2\pi = 9.75$ GHz.

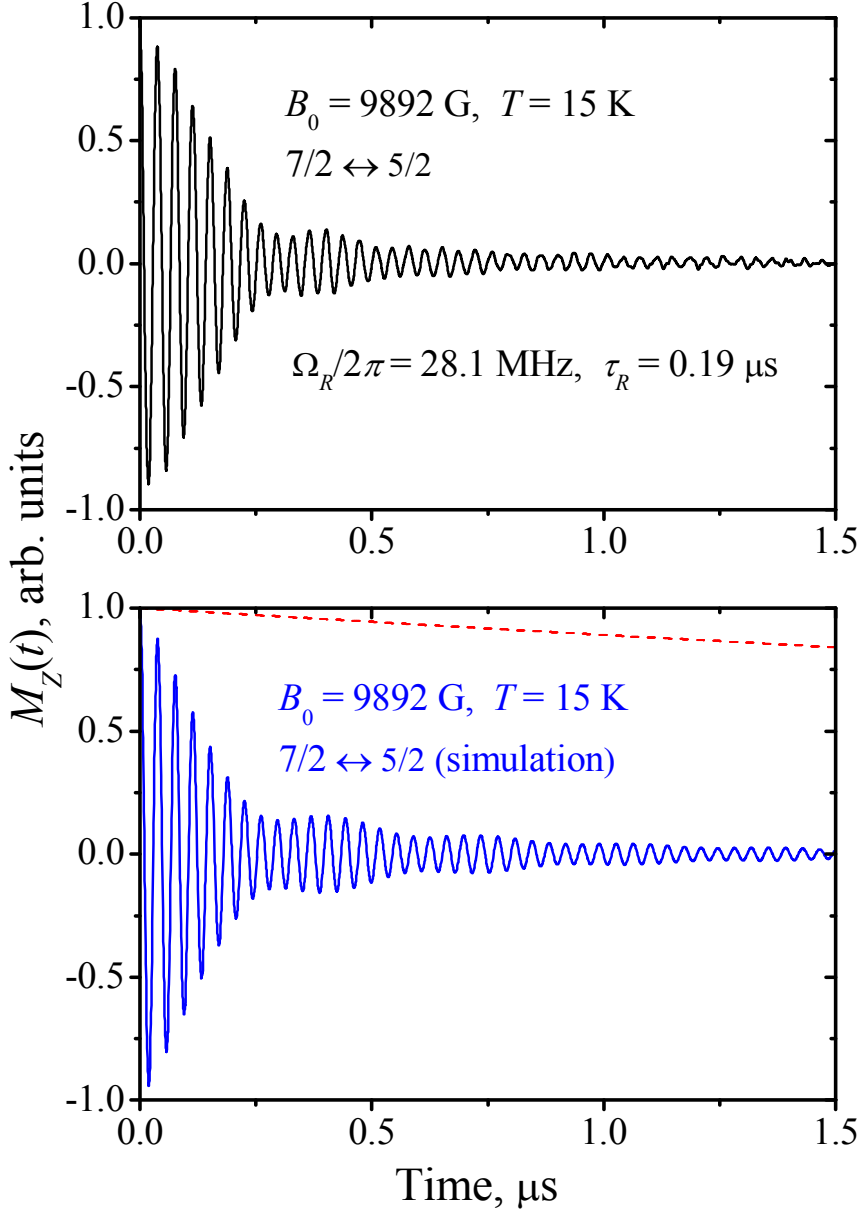

FIG. 8. Rabi oscillations of the $7/2 \leftrightarrow 5/2$ transition at 9892 G, $T = 15$ K. The experimental data and the calculations of the average magnetic moment according to the Eq. (8) are shown by the black (top) and blue (bottom) solid lines, respectively. The red dashed line represents the calculated dipolar contribution to the amplitude damping $\sim \exp(-\Gamma_d t)$.

Since only two out of 8 electronic levels were bound by the mw field at certain $B_0$, and $T_1$ times were several degrees of magnitude longer than the Rabi period, one could ignore the nonresonant $M \rightarrow M'$ transitions and use an effective spin-1/2 approach. The experimental data were successfully simulated by the following expression [33,34]:

$$M_Z(t) \sim \int_V d\mathbf{r} \int g(\omega) d\omega \frac{\cos\left(\sqrt{\Omega_R^2(\mathbf{r}) + (\omega - \omega_0)^2} \cdot t\right)}{\Omega_R^2(\mathbf{r}) + (\omega - \omega_0)^2} \times \exp(-\Gamma_d t). \qquad (8)$$

Here $\sqrt{\Omega_R^2(r)+(\omega-\omega_0)^2}$ is the nutation frequency of the spin packet detuned by $\omega-\omega_0$ from the resonant frequency, $\Omega_R(r) = g\mu_B B_1(r)\sqrt{S(S+1)-M(M+1)}/2\hbar$ corresponds to the local Rabi frequency corresponding to the $M \leftrightarrow M+1$ transition of the $S$-spin particle situated at the site $r$ within the crystal sample, $r=0$ corresponds to the field antinode, and $\Omega_R = \Omega_R(r=0)$ is related to the most "coherent" part of the spin ensemble. The integration over the sample volume $V$ and over different spin packets $\omega$ within the resonance line was performed numerically. The measured distribution of the mw field amplitude along the resonator axis $B_1(x)$ was well approximated by the generalized Gaussian distribution (Fig. 3). We assumed analogous distribution $B_1(y,z) \sim \exp\left[-(\rho/c)^d\right]$ in the perpendicular plane ($\rho^2 = y^2 + z^2$), where the plausible parameters $c = 2.5$ mm and $d = 2.4$ gave the best fit of the experimental data. That corresponded to the maximum inhomogeneity of $\sim 15$ % within the sample volume, i.e. $0.85\Omega_R \leq \Omega_R(r) \leq \Omega_R$. The amplitude of ROs had polynomial decay $\sim \left[1+(t/\tau_R)\right]^{-\kappa}$ associated with the $B_1$-type dephasing of the spin packets, and was modulated due to the constructive/destructive interference of the nutations from the different areas of the sample [33,34].

The exponential factor in Eq. (8) accounts for the magnetic dipole interactions between the spins. The mechanism of the dipolar coupling-induced spin relaxation in the presence of the driving mw field was studied previously [34,35]. It was shown that only those spins that were driven by the mw field contributed effectively to the local magnetic field through $S_x^j S_x^k$ terms of the dipolar coupling, and thus influenced the decay time. The corresponding decay rate is proportional to the dipolar half-width multiplied by the population factor $\gamma_T$, cf. Eq. (3):

$$\Gamma_d = \lambda(\sigma/\Omega_R) \cdot \gamma_T \Delta\omega_d^{(1/2)}, \qquad (9)$$

where the factor $0 < \lambda(\sigma/\Omega_R) \leq 1/2$ depends on the portion of the spin packets contributing to the random local field. In particular, $\lambda = 1/2$ corresponds to the case of high mw field, or narrow resonance lines ($\sigma \ll \Omega_R$), when almost all spin packets are driven by the microwaves. The data shown in Figs. 7 and 8 corresponded to the latter case. Generally, for a Lorentzian distribution of the resonance frequencies, one can calculate $\lambda(\nu)$ as in Ref. [34]:

$$\lambda(\nu) = \begin{cases} \dfrac{\arccos\nu}{\pi\sqrt{1-\nu^2}}, & \nu < 1, \\ \dfrac{\ln(\nu+\sqrt{\nu^2-1})}{\pi\sqrt{\nu^2-1}}, & \nu > 1. \end{cases} \qquad (10)$$

The results of the calculations by the Eq. (8) shown on the bottom part of Fig. 8 agree well with the experimental data. A red dashed line indicates the dipolar contribution with the decay time $\Gamma_d^{-1} = 11$ μs, which is nearly two orders of magnitude greater than the overall decay time $\tau_R = 0.19$ μs.

In order to study $\tau_R(\Omega_R)$ dependence and test our model, we measured ROs of the high-field ($7/2 \leftrightarrow 5/2$) transition at varying levels of the mw field attenuation and temperature 6 K. The corresponding range of Rabi frequencies $0.97 \leq \Omega_R/2\pi \leq 32.5$ MHz covered both the low-field ($\sigma > \Omega_R$) and the high-field ($\sigma < \Omega_R$) regions, with $\sigma/2\pi = 7$ MHz. A change in transient dynamics of the spin ensemble while switching from partial to total line excitation is shown in Fig. 9, while the resulting $\tau_R^{-1}(\Omega_R)$ dependence is depicted in Fig. 10.

In the first case, when $\sigma > \Omega_R$, there are a number of spin packets detuned sufficiently from resonance, with the nutation frequencies $\sqrt{\Omega_R^2 + (\omega - \omega_0)^2} \neq \Omega_R$. An integration over $\omega$ in the Eq. (8) sums up all these contributions and produces nonexponential decay $\sim \left(1 + (\pi\Omega_R t)^2\right)^{-1/4}$ associated with the spin dephasing [34,36]. This decay is significant during the first few oscillations, when the contributions from the out-of-resonance spin packets rapidly fade away. An interplay between this contribution (that we will further call "$\omega$-type dephasing") and the $B_1$-type dephasing gives the $\tau_R^{-1}(\Omega_R)$ dependence nonlinear in $\Omega_R$, see the left part of Fig. 10. In the second case, when $\sigma < \Omega_R$, one can neglect the $\omega$-type dephasing. The Rabi decay rate is linear in $\Omega_R$, which indicates the dominant role of the $B_1$-type contribution. At the highest $B_1$ field, we observed up to 100 Rabi half-periods (Fig. 9f).

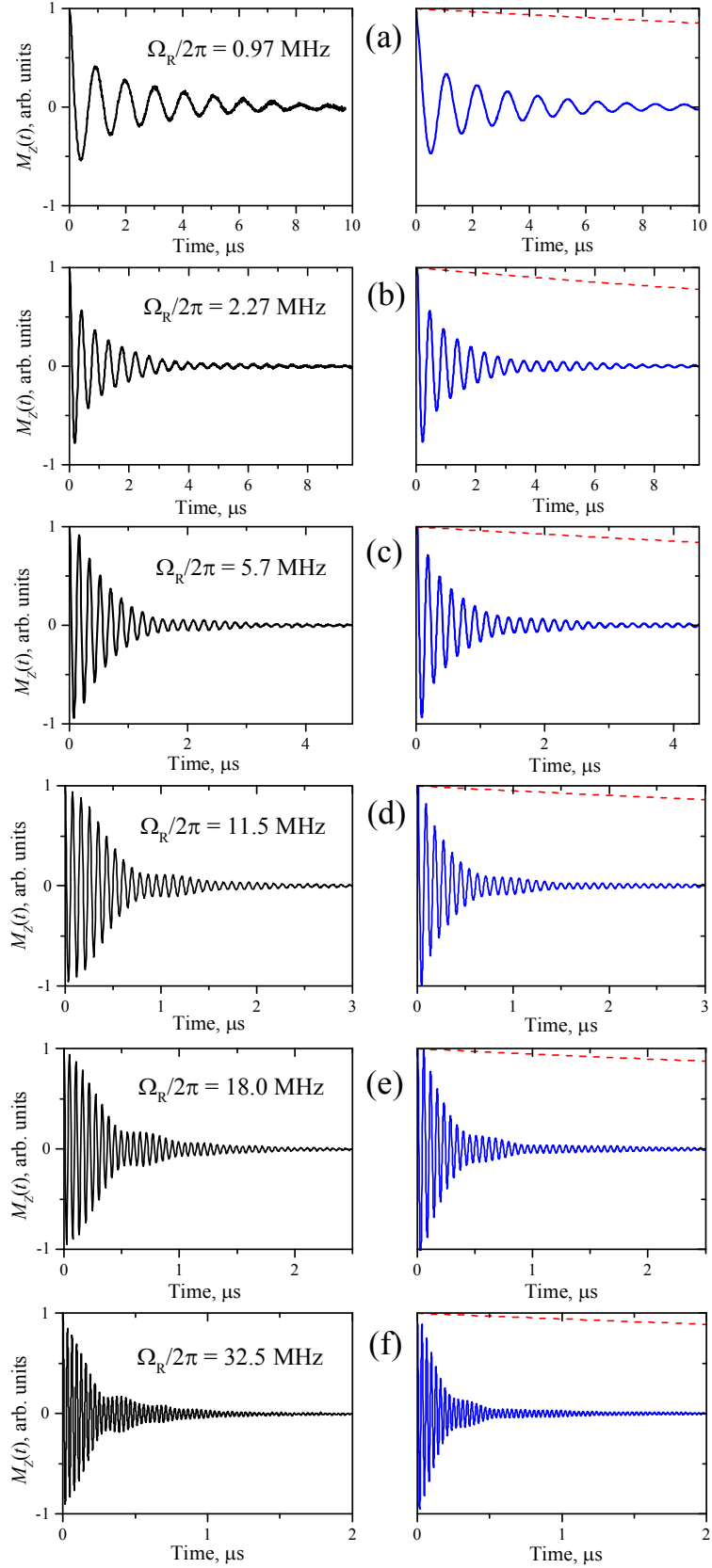

FIG. 9. Rabi oscillations of the $7/2 \leftrightarrow 5/2$ transition at varying levels of the mw field attenuation. $T = 6$ K. The experimental data and the calculations of the average magnetic moment according to the Eq. (8) are shown by the black (left) and blue (right) solid lines, respectively. The red dashed lines on the right figures represent the calculated dipolar contributions to the amplitude damping ~ $\exp(-\Gamma_d t)$.

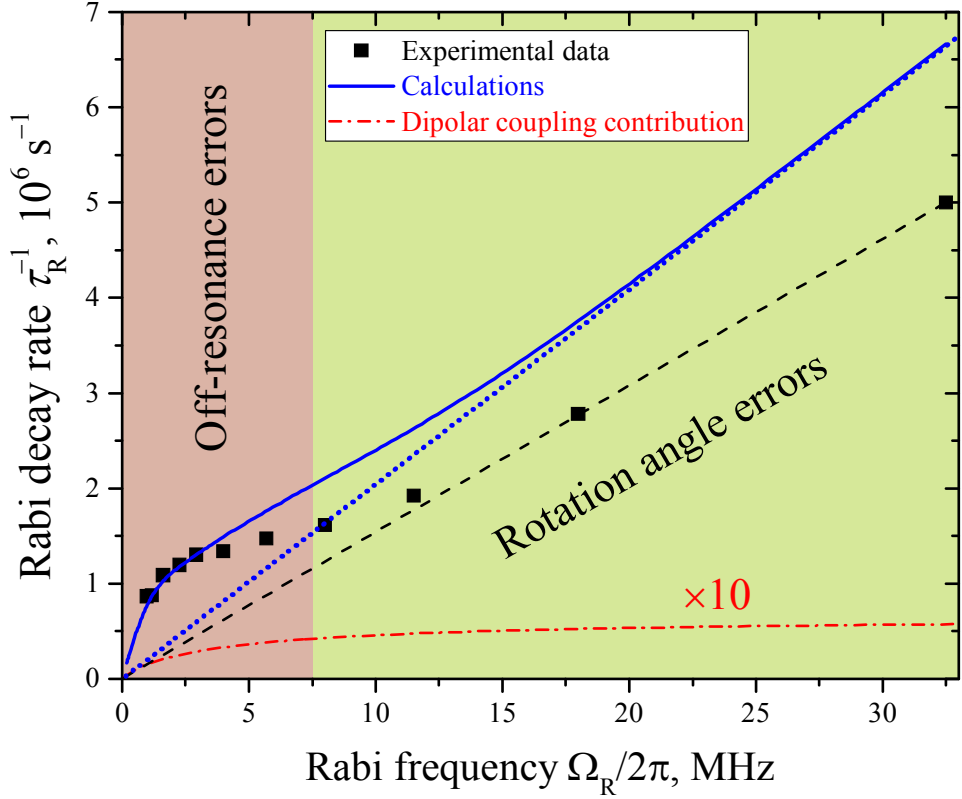

FIG. 10. Experimental (black squares) and calculated (blue solid line) inverse Rabi times vs. Rabi frequency. Black dashed and blue dotted lines represent linear approximations $\tau_R^{-1} = \beta\Omega_R$ of the corresponding data at $\Omega_R \gg \sigma$, with $\beta = 0.024$ and $\beta = 0.032$, respectively. Red dash-dotted line represents the dipolar decay rate $\Gamma_d(\Omega_R)$ calculated by Eq. (9) and multiplied by a factor of 10. The low- and high-field areas shown in pink ($\Omega_R < \sigma$) and green ($\Omega_R > \sigma$) correspond to regions where the dominant contribution comes from either $\omega$-type dephasing (giving rise to off-resonance errors) or $B_1$-type dephasing (rotation angle errors).

In magnetic resonance quantum computing, the $\omega$-type and $B_1$-type dephasings can be related to off-resonance and rotation angle errors, respectively [37]. The two regions where either of these errors dominates are shown in Fig. 10. There, one can clearly see the transition from the nonlinear behavior at $\Omega_R < \sigma$ to the linear one at higher fields: $\tau_R^{-1}(\Omega_R \gg \sigma) \simeq \beta\Omega_R$ with $\beta = 0.024$. The calculated $\tau_R^{-1}(\Omega_R)$ dependence qualitatively agrees with the experimental data. However, while the decay profiles and the Rabi amplitude modulation are represented correctly (see Figs. 8, 9), our estimated $\beta$ value is ~ 30% higher. This discrepancy possibly originates from the oversimplified modelling of the $B_1(y,z)$ field dependence, actual nonrectangular shape of the crystal sample, and small inaccuracy in positioning the sample with respect to the center of the resonator.

As shown in Fig. 10, the role of the dipolar coupling was rather small in comparison with the spin dephasing in the whole studied range of $\Omega_R$. Due to smaller thermal occupation of the upper 7/2 and 5/2 levels involved in the studied transition at 6 K, the calculated dipolar relaxation time $\Gamma_d^{-1}$ reached 17 μs at the highest mw field strength. This value provides an upper limit for the Rabi damping time in the case when the mw field inhomogeneity is suppressed, which would correspond to ~ $10^3$ one-qubit NOT operations at $\Omega_R/2\pi = 32.5$ MHz. This can be achieved by using smaller samples (however, at the sacrifice of the signal-to-noise ratio), or utilizing the rotation angle and off-resonance correction schemes like BB1 and CORPSE composite pulse sequences [37-40].

## IV. CONCLUSIONS

In the present article, coherent manipulations of the quantum states belonging to the spin-7/2 trivalent gadolinium ion hosted in CaWO$_4$ single crystal are implemented. Coherence times and Rabi oscillations were acquired at different lines of fine and hyperfine structure of the Gd$^{3+}$ ion hosted in CaWO$_4$ single crystal. Spin-lattice relaxation times up to 10 ms and phase memory times up to 25 μs were recorded in the temperature interval 6-150 K. The Rabi damping times and the phase memory times were interpreted in the framework of a model that takes into account the magnetic dipole interactions between the ions and, in the case of the Rabi damping, the spin dephasing associated with the intrinsic inhomogeneity of the microwave field inside the resonator and the inhomogeneous broadening of the resonance line.

The observed rich level structure of the S-state Gd$^{3+}$ ion in CaWO$_4$ crystal provides the opportunity to tune the microwave frequency, the magnetic field and the sample orientation in order to apply appropriate quantum computation scheme. We experimentally demonstrated the possibility to perform up to 100 simple one-qubit operations on various X-band EPR transitions of the lowest electronic $^8S_{7/2}$ multiplet (this figure can be potentially increased to ~ $10^3$ using the off-resonance and the rotation angle error correction schemes). In comparison with the previously studied spin-5/2 systems [16-18], the present system offers somewhat larger Hilbert space of $8 = 2^3$ addressable quantum states, meaning that each Gd$^{3+}$ ion can represent three effective qubits. Moreover, $^{155}$Gd$^{3+}$ and $^{157}$Gd$^{3+}$ ions with nuclear spin $I = 3/2$ are potential hybrid 5-qubit systems, whose states can be accessed, for instance, via pulsed ENDOR. Nearby nuclear spins, if present, can lead to resolved superhyperfine structure, as was recently observed for $^{19}$F nuclei in LiYF$_4$:Gd$^{3+}$ crystal [21], thus giving additional quantum states for the hybrid quantum computing.

## ACKNOWLEDGMENTS

The work was partially supported by the program of competitive growth of Kazan Federal University. EIB was funded by the Russian President Fellowship for Young Scientists (stipend no. 924.2015.5). The authors thank G. V. Mamin for his assistance in experimental work.